\documentstyle [preprint,prb,aps]{revtex}

\begin{document}
\draft
\title{n-Alkyl Thiol Head Group Interactions with the Au(111) Surface}
\author{Y. Yourdshahyan,  H.K. Zhang, and A.M. Rappe}  
\address{Department of Chemistry and Laboratory for Research on the 
Structure of Matter, \\
 University of Pennsylvania, Philadelphia, PA 19104-6323 }

\maketitle

\begin{abstract} 
{ State-of-the-art first principles calculations based on density  
functional theory were performed on  CH$_3$(CH$_2$)$_{n-1}$S-Au(111) 
systems. We show that the adsorption site of  methylthiolate at low 
coverage on the Au(111) surface is the fcc site, not the hcp site  
as  has been recently reported. Further, we report results for chain 
length dependency and the electronic structure of the system. } 
\vspace{0.5cm}
\end{abstract}

The interaction between organic materials and solid surfaces has 
been extensively studied because of the broad range of industrial 
applications.\cite{Ulm,Mab,DiMi,Chai} Self-assembled monolayers 
(SAMs) hold special interest, because the presence of the thiol 
group  greatly strengthens the molecule-surface interactions,  
inducing order in the layer. SAMs have important potential applications 
in industry, such as sensors, transducers, detectors, packaging 
and insulating layers for integrated circuits, functionalization 
of surfaces, thin coatings for electrodes, and  corrosion inhibition. 

The long-chain alkane thiols [CH$_3$(CH$_2$)$_{n-1}$SH, or C$_n$] 
form SAMs  on the Au(111) surface. Their simplicity, highly ordered 
structures, and chemical stability make these systems ideal for study 
with a variety of techniques including  atomic force microscopy,\cite{Tama,Xu}  
infrared spectroscopy,\cite{Bain,Nuzz} high-resolution electron-energy-loss 
spectroscopy,\cite{Klut} grazing X-ray diffraction,\cite{Fen3} 
scanning tunneling microscopy (STM),\cite{Poi3,Cam,Arce,Yama,Kond,Port,Scho} 
scanning probe microscopy,\cite{Alve} low-energy electron 
diffraction (LEED),\cite{Cam1,Dubo,Chid1} He atom diffraction,\cite{Cam2,Chid}  
and theory.\cite{Bahi,Gerd,Mar,Haut,Pert,Sell,Bea}
Despite extensive studies of this system, their are many controversies
regarding its structure. 

Two competing structural models, the ``standard model'' and 
the ``sulfur-pairing model'', have considerable support. In 
the standard model, alkanethiol molecules exhibit a hexagonal 
($\sqrt{3}\times\sqrt{3}$)R30$^\circ$ lattice.
Alkanethiolates occupy three-fold hollow sites of the Au 
(111) surface with S-S spacings of 4.99 \AA,  and the molecular 
axes are tilted by 30$^\circ$-35$^\circ$ with respect to the surface 
normal.\cite{Arce,Yama,Kond,Port,Scho,Alve,Cam1,Dubo,Chid1,Cam2,Chid,Sell,Bea,Bieb,Zhor,Nuzz1,Carr} 
The sulfur-pairing model involves a $c$(4 $\times$ 2) 
superlattice of the hexagonal lattice, where alkanethiolates 
form sulfur head group dimers with S-S spacings of 2.2 
\AA.\cite{Klut,Fen3,Poi3,Cam,Nis,Bahi,Gerd} Furthermore, other 
structural models have been reported, including  $n\times\sqrt{3}$ 
unit cells ($8\leq n \leq10$).\cite{Cam4,Wan,Poi-1,Kang} There is 
even controversy among proponents of the sulfur pairing model, 
with some researchers suggesting top and hollow-bridge positions 
for the sulfurs, while others propose a hollow and hollow-bridge dimer. 

In order to understand the structure fully one must include the effect
of coverage, chain length, and temperature. However, because the 
differences between these structures center around the thiol head-group 
locations, theoretical investigation of the thiol-gold chemisorption
bond is vital. Since interactions with the metal surface appear to
dominate the molecular arrangement of SAM systems, the interaction of 
a methylthiolate (CH$_3$S) with the Au (111) surface presents a prototype 
system to study this entire class of systems.

Both LEED\cite{Dubo} and ultrahigh-vacuum cryogenic STM\cite{Kond} 
studies show a ($\sqrt{3}\times\sqrt{3}$)R30$^\circ$  pattern of CH$_3$S 
monolayers at room temperature, with head-groups at three-fold hollow
sites, while at 110 K a disordered hexagonal ($3\times4$) lattice has 
been reported.\cite{Kond} The S-S distances in the latter study (3.3 \AA ) 
are much longer than the 2.2 \AA\  and much shorter than 4.99 \AA\ 
from the sulfur-pairing and standard models. Two theoretical studies 
have concluded that the adsorption site for a single thiol molecule on 
the gold (111) surface is the hcp hollow site.\cite{Sell,Bea} These 
results also support the standard model. However, despite the overall 
agreement of the results, there are discrepancies regarding the S-Au 
distance and the tilt of the molecule in these two independent studies. 

The main objective of our study is to identify the adsorption site 
for thiols on gold (111), and to explain the structural properties. 
We have performed first-principles calculations based on 
density-functional theory (DFT)\cite{kohn} to study the adsorption 
of methylthiolate on the Au (111) surface. Our results show that a 
single CH$_3$S molecule adsorbed on the Au (111) surface chemisorbs 
on the fcc site. We also find that further increase of the hydrocarbon 
chain does not affect the adsorption site. All of the first-principles 
calculations have been performed with the plane-wave pseudopotential 
code ``dacapo'' \cite{daca}.

In our  DFT calculations, the wave functions are expanded 
in a plane-wave basis set, the electron-ion interactions 
are described by ultra-soft pseudopotentials (USPP) \cite{Van}, 
and the generalized gradient approximation (GGA) \cite{PBE} 
for the exchange-correlation functional has been used. The 
Kohn-Sham equations are solved self-consistently, using a Pulay  
density-mixing  scheme \cite{kres} to update the electronic density 
between iterations. The occupation numbers are updated using a 
recently developed technique based on minimization of the 
free-energy functional \cite{Beng}. A finite electronic temperature 
is used, in order to reduce the number of $k$ points needed, 
and all total energies are then extrapolated to zero electronic 
temperature. For the optimization of atomic structure, a damped 
molecular-dynamics (DMD) method has been used.

When using the slab geometry in the plane wave pseudopotential 
method, errors most often arise from the pseudopotential or from 
lack of convergence with respect to the  number of $k$-points, 
the cutoff energy, and the system size. When generating pseudopotentials it is 
of great importance to preserve the eigenvalues for all relevant 
atomic configurations, not just in the reference configuration. 
However, in order to have correct lattice constants for metals it 
is also of importance to preserve the charge density in the tail 
region for all relevant atomic configurations \cite{Illy}. For H 
(1s$^1$), C (2s$^2$2p$^2$), S (3s$^2$3p$^{3.5}$3d$^{0.5}$), and Au 
(5d$^{9.5}$6s$^1$6p$^{0.5}$) the USPPs were generated using cutoff 
radii (r$_c$) of  0.60, 1.24, 1.45, and  2.00 bohr, respectively, 
and the maximum transferability errors were less than 2 mRy.  All
potentials are tested and compared to experiment and to all-electron
calculation. 

We find that although the $3d$ orbitals of S are nearly unpopulated 
in simple molecules, such as S$_2$, the $d$  channel has great impact 
on the behavior of the pseudopotential. The 3$d$ orbitals give S 
enhanced polarizability  and enable a great variety of bonding 
configurations through  hybridization. A comparison between two 
USPPs for S with and without the $d$-projector (transferability errors 
less than 2 mRy in both potentials) show binding energies 
(E$_{b}$ of S$_2$) of 2.51 and 3.08 eV/atom (all-electron and
experimental values are 2.49 eV/atom \cite{Xi} and 2.37 eV/atom 
\cite{exp}), respectively. 

In order to reduce calculation errors due to finite slab thickness, 
different slab geometries (3-8 layers thick) have been tested using 
different numbers of $k$-points and vacuum layers (see Fig.~\ref{fig:conv}).  A 
slab consisting of 6+7 Au+vacuum layers, a cutoff energy of 30 Ry, 
and  a 4$\times$4$\times$1 grid of special $k$-points is found to 
give converged results (convergence within a few meV). Table I shows 
the equilibrium parameters for the  clean Au (111) 
surface and the free CH$_3$S and CH$_3$SH  molecules 
compared to experiment and other calculations. 

The first step toward a detailed understanding of the SAM 
structure is to investigate the adsorption site of the simplest
C$_n$ (CH$_3$S) on the surface. Therefore, we calculated the 
total energy of a single CH$_3$S molecule in a (2$\times$2) 
surface unit cell, for a coverage of 0.25.
As shown in Fig.~\ref{fig:stru}a, the thiolate was moved 
from top to fcc, then over the bridge site to the hcp site 
using different molecular orientations.  The supercell  
consisted of  6 Au(111) layers and 7 layers of vacuum. 
The three topmost Au layers and the molecule were optimized 
using the DMD method until the total force for the system was 
less than 0.01 eV/\AA. Figure \ref{fig:bar} shows calculated 
potential energy surfaces for two different molecular orientations
along the diffusion path. The dashed curve is for the orientation
used in Ref.~\onlinecite{Sell} and the solid line is the minimum-energy 
path. Our results show that the staggered configuration is 
preferred over the configuration used in previous studies.
The results of surface buckling and other optimization effects 
for the preferred orientation (solid line) are summarized in 
Table II and shown schematically in Fig.~\ref{fig:stru}b. 

The calculated chemisorption energy difference between fcc and hcp,
$\Delta$E$_{\rm hcp-fcc}$ = 0.10 eV, shows that a single CH$_3$S
molecule adsorbed on the gold surface prefers the fcc site over hcp.
This is in contrast with results of Refs.\ \onlinecite{Sell} and
\onlinecite{Bea}, which are based on cluster calculations and
classical MD. In the cluster calculations, only two gold layers were
used. As a result, the surface energy is far from converged, and the
effect of gold atoms in the third layer is not included. As can be
seen from Fig.~\ref{fig:conv}, the gold (111) surface energy is
converged only when six or more gold layers are used. Furthermore, the
energy difference between having the molecule at the fcc or hcp sites,
($\Delta$E$_{\rm hcp-fcc}$) increases from 0.04 to 0.10 eV as the slab
thickness is increased from four to six gold layers.  This indicates
the importance of slab size and suggests that the use of a two-layer
gold slab is insufficient for accurate results.

Regarding the effect of hydrocarbon chain length, we performed
calculations using two and three carbons in the chain. The results
show no changes in the preferred adsorption site (fcc), and
$\Delta$E$_{\rm hcp-fcc}$ is 0.15 eV and 0.18 eV for two- and
three-carbon thiols, respectively. These calculations indicate that
the preference for fcc becomes stronger with increasing number of
carbons in the chain.

The induced charge density (Fig.\ 4) shows an increase of charge between
the Au and S atoms, when a thiol molecule is adsorbed on the fcc site.  
The metallic electrons (via Pauli repulsion) shift this region of 
enhancement, so that it is farther from the surface than expected. 

In summary, we have presented state of the art DFT calculations of
thiol molecules adsorbed on the Au (111) surface at low coverage.
This study demonstrates that the fcc site is the preferred location
for a single thiol, and the electronic changes which accompany
chemisorption are elucidated.  Furthermore, the importance of having a
minimum of six gold layers in the model, and of including $d$ orbitals
in the sulfur pseudopotential were highlighted.

The authors wish to acknowledge G. Scoles for his insightful 
comments concerning this subject.  This work was supported by 
NSF grant DMR 97-02514 and the Air Force Office of Scientific 
Research, Air Force Materiel Command, USAF, under grant number 
F49620-00-1-0170. AMR would like to thank the Alfred P. Sloan 
Foundation for support. Computational support was provided by 
the National Center for Supercomputing Applications and the 
San Diego Supercomputer Center.




\noindent
\begin{table}
\caption{Structural and energetic results for the clean gold (111) 
surface (slab consists of 6 Au and 7 vacuum layers) and the CH$_3$S 
and  CH$_3$SH molecules. Comparison of calculated results for the 
interlayer relaxations $\Delta d_{12}$, $\Delta d_{23}$, and 
$\Delta d_{34}$ (1 = top layer) of the Au(111) surface.  
The presented distances are in \AA\ ,  and the surface energy (E$_s$) 
is given in eV/\AA$^2$. }
\vspace{0.25cm}
\label{tab:Au}
\begin{tabular}{l c c c c }
\multicolumn{5}{l}{Clean Au (111) surface:} \\
   & $\Delta d_{12}$ (\%)& $\Delta d_{23}$ (\%)& 
$\Delta d_{34}$ (\%) & E$_s$ \\
This work & 0.97 & --0.48 & 0.07 & 0.101 \\
Calc. \cite{Cox} & \hspace{-0.15cm}--0.24 & \hspace{0.15cm}0.05 & 
0.04 & 0.084 \\
Expt. \cite{Van1} & 0.00 & & &0.096 \\
\hline
\multicolumn{5}{l}{Optimized parameters for CH$_3$S:}\\
& $r$(CS)  & $r$(CH$_a$)  & $r$(CH$_b$) & $r$(CH$_c$)  \\
This work&1.789 & 1.070 & 1.068 & 1.068 \\
Calc. \cite{See}& 1.799 & 1.095 & 1.091 & 1.091  \\
& $\theta$(SCH$_a$)&$\theta$(SCH$_b$)&$\theta$(SCH$_c$)& 
$\phi$(H$_a$SCH$_b$) \\
This work& 109.8$^\circ$  & 110.7$^\circ$  & 110.7$^\circ$  & 116.3$^\circ$ \\
Calc. \cite{See}&107.0$^\circ$ & 111.6$^\circ$ & 111.6$^\circ$ & 
118.0$^\circ$ \\
\hline
\multicolumn{5}{l}{Optimized parameters for CH$_3$SH:}\\
& $r$(CS)  & $r$(CH) & $r$(SH$_d$) & $\theta$(CSH$_d$)\\
This work&1.827 & 1.08 & 1.36 & 97.2$^\circ$  \\
Expt. \cite{See1}  & 1.819& 1.09 & 1.34 &96.5$^\circ$
\end{tabular}
\end{table}


\begin{table}
\caption{Calculated parameters for CH$_3$S interaction with the Au (111) 
surface. The parameters $d_{\rm C-S}$, $d_{\rm S-Au}$, and $d_{\rm x,y}$ 
are the distances between the C and S atoms, between the  S and the center 
of mass (CM) of the topmost Au layer, and the interlayer separation 
between the CMs of two adjacent layers. $\delta_{xa}$, 
$\delta_{xb}$, and $\delta_{xc}$ give the surface buckling. }
\label{tab:Au1}
\begin{tabular}{l c c c c}
			& fcc     & hcp     & bri & top\\
$\Delta$E$_{\rm chem}$ (eV)	& 0.00   & 0.10   & 0.40  & 0.95\\
$d_{\rm S-Au}$ (\AA )             & 1.788   & 1.831   & 1.990  & 2.493\\
$d_{12}$      (\AA )      	& 2.399   & 2.398   & 2.430  & 2.410\\
$d_{23}$     (\AA )       	& 2.386   & 2.382   & 2.398  & 2.379\\
$d_{34}$  	(\AA )    	& 2.393   & 2.389   & 2.401  & 2.398\\
$\delta_{1a}$ 	(\AA )   	& 0.001   & 0.005   & 0.008  & 0.015\\
$\delta_{1b}$ 	 (\AA )    	& 0.004   & 0.006   & 0.007  & 0.011\\
$\delta_{1c}$ 	 (\AA )    	& --0.003 & --0.007 & --0.008  & --0.021\\
$\delta_{2a}$ 	 (\AA )    	& 0.000   & 0.000   & 0.004  & 0.010\\
$\delta_{2b}$ 	 (\AA )  	& 0.002   & 0.002   & 0.005 & 0.009\\
$\delta_{2c}$ 	 (\AA )  	& --0.003 & --0.003 & --0.004  & --0.003\\
$r$(CS)          (deg.)    	& 1.848   & 1.846   & 1.840  & 1.835\\
$r$(CH$_a$) 	  (deg.)	& 1.105   & 1.105   & 1.106  & 1.106\\
$r$(CH$_{b,c}$)   (deg.)	& 1.104   & 1.104   & 1.105  & 1.106\\
$\theta$(SCH$_a$)  (deg.)	& 108.6   & 108.7   & 108.9  & 109.1\\ 
$\theta$(SCH$_{b,c}$)  (deg.)   & 108.5   & 108.5   & 108.8  & 108.9\\

\end{tabular}
\end{table}


\begin{figure}
\caption{The Au(111) surface energy in eV/\AA$^2$ as a function 
of Au layers for different numbers of $k$-points and vacuum thickness. All 
calculations are performed using a 30 Ry cutoff energy. }
\label{fig:conv}
\end{figure}

\begin{figure}
\vspace{0.5cm}
\caption{The adsorption energy for a CH$_3$S molecule  on the Au(111) 
surface along the diffusion path shown in Figure 3. Small figures show 
top views for molecular orientation at each site. Solid and dashed
lines represent energies for orientations shown below and above 
the curve, respectively.}
\label{fig:bar}
\end{figure}

\begin{figure}
\vspace{0.5cm}
\caption{Schematic presentation of the interaction of CH$_3$S with
the Au(111) surface. T, H, B, and F denote top, hcp, bridge, and fcc 
sites, respectively. d$_{xy}$ is the interlayer distances between 
layers $x$ and $y$, where the top layer is denoted 1.  
$\delta_{xa}$ and $\delta_{xb}$ represent buckling for nearest and  
$\delta_{xc}$ for the  next-nearest neighbors of the S atom. }
\label{fig:stru}
\end{figure}

\begin{figure}
\vspace{0.5cm}
\caption{Calculated induced and total charge density when the
methylthiolate is adsorbed on the fcc site. 3-D isosurface (a) and a
slice parallel to the surface between S and Au (c) of the induced
charge density, and a slice of the total charge density (b) are
presented.}
\label{fig:bar}
\end{figure}

\end{document}